\documentclass[superscriptaddress,showpacs,twocolumn,pre,floatfix]{revtex4-1}
\usepackage{graphicx}
\usepackage{color}
\usepackage{appendix}
\usepackage{amsmath}
\usepackage{amssymb}
\usepackage{subfigure}
\usepackage{epsfig}

\newcommand{\xh}{\hat{x}}
\newcommand{\yh}{\hat{y}}
\newcommand{\etab}{\boldsymbol{\eta}}
\newcommand{\epsilonb}{\boldsymbol{\epsilon}}

\newcommand{\U}{\boldsymbol{U}}

\newcommand{\B}{\boldsymbol{B}}
\newcommand{\E}{\boldsymbol{E}}
\newcommand{\one}{\boldsymbol{1}}
\newcommand{\D}{\boldsymbol{D}}
\newcommand{\C}{\boldsymbol{C}}
\newcommand{\rr}{\boldsymbol{r}}
\newcommand{\kap}{\boldsymbol{\kappa}}
\newcommand{\sig}{\boldsymbol{\sigma}}
\newcommand{\gams}{\dot{\gamma}_{s}}
\newcommand{\gamc}{\dot{\gamma}_{c}}
\newcommand{\gamone}{\dot{\gamma}_1}
\newcommand{\gamtwo}{\dot{\gamma}_2}
\newcommand{\eff}{\varepsilon_{\text{eff}}}

\begin{document}
\title{
Three Dimensional Flow of Colloidal Glasses  
}

\date{\today}
\author{T.F.F. Farage}
\email{thomas.farage@unifr.ch}
\affiliation{Department of Physics, University of Fribourg, CH-1700 Fribourg, Switzerland}
\author{J.M. Brader}
\email{joseph.brader@unifr.ch}
\begin{abstract}
Recent experiments performed on a variety of soft glassy materials  
have demonstrated that any imposed shear flow serves to simultaneously fluidize these systems in all
spatial directions [Ovarlez \textit{et al.} (2010)]. 
When probed with a second shear flow, the viscous response of the experimental system is determined by the
rate 
of the primary, fluidizing flow.  
Motivated by these findings, we employ a recently developed schematic mode-coupling theory 
[Brader \textit{et al.} (2009)] to investigate the three dimensional flow of a colloidal glass,
subject
to a combination of 
simple shear and uniaxial compression. 
Despite differences in the specific choice of superposed flow, the flow curves obtained show good qualitative 
agreement with the experimental findings and recover 
the observed power law describing the decay of the scaled viscosity as a function of the dominant
rate. 
We then proceed to perform a more formal analysis of our constitutive equation
for different kind of `mixed'
flows 
consisting of a dominant primary flow subject to a weaker perturbing flow. 
Our study provides further evidence that the theory of Brader \textit{et al.} (2009) reliably
describes
the  
dynamic arrest and mechanical fluidization of dense particulate suspensions. 
\end{abstract}


\maketitle
\section{Introduction}
Colloidal dispersions display a broad range of nontrivial rheological response to externally 
applied flow.  
Even the simplest systems of purely repulsive spherical colloids exhibit a rate dependent 
viscosity in steady state flows, yielding and complex time-dependent phenomena, such as 
thixotropy and ageing [Brader (2010), Mewis and Wagner (2009)].  
Understanding the emergence of these collective dynamical phenomena from the underlying 
interparticle interactions poses a challenge to nonequilibrium statistical mechanics and 
the fundamental mechanisms involved are only beginning to be understood. 
Theoretical advances have largely been made hand-in-hand with improved simulation techniques
[Banchio and Brady (2003)]
and modern experimental developments, combining confocal microscopy or magnetic resonance imaging 
with classical rheological measurements [Besseling \textit{et al.} (2010), Frank \textit{et al.}
(2003)].

Despite considerable progress, a comprehensive constitutive theory, capable of capturing the full 
range of response, remains to be found. 
Existing approaches are tailored to capture the physics of interest within particular ranges of 
the system parameters (e.g. density, temperature) but fail to provide the desired global framework. 
Moreover, the vast majority of studies have concentrated on the specific, albeit important, 
case of simple shear flow. 
Such scalar constitutive theories, relating the shear stress to the shear strain and/or
strain-rate, 
provide important information regarding the competition of timescales underlying the rheological  
response, but do not acknowledge the true three dimensional character of experimental flows. 
Tensorial constitutive equations have long been a staple of continuum rheology 
(such as the Giesekus or Oldroyd models [Bird \textit{et al.} (1987), Larson (1988)]) and enable
e.g. normal forces and 
secondary flows to be addressed in realistic curvilinear experimental geometries. 

The first steps towards a unified, three dimensional description of colloid rheology have been 
provided by recent extensions of the quiescent mode-coupling theory (MCT) to treat dense 
dispersions under flow [Brader \textit{et al.} (2008)]. 
These developments are built upon earlier studies focused on simple shear
[Brader \textit{et al.} (2007), Fuchs (2009), Fuchs and Cates (2002), Fuchs and Cates (2009)]
and capture the competition between slow structural relaxation and external driving,  
thus enabling one of the most challenging aspects of colloid rheology to be addressed: 
the flow response of dynamically arrested glass and gel states. 
Given the equilibrium static structure factor as input (available from either simulation or 
liquid state theory [Brader (2006)]), the deviatoric stress tensor $\sig(t)$ may be determined for
any given velocity gradient tensor $\kap(t)$.
However, implementation of the theory has been hindered by the numerical resources required to
accurately integrate 
fully anisotropic dynamics over timescales of physical interest (although progress has been made for 
two dimensional systems [Henrich \textit{et al.} (2009), Kr\"uger \textit{et al.} (2011)]). 
In [Brader \textit{et al.} (2009)] a simplified `schematic' constitutive model was proposed, which
aims
to capture the essential 
physics of the wavector dependent theory, while remaining numerically tractable. 
Applications so far have been to steady-state flows, step strain and dynamic yielding [Brader
\textit{et
al.} (2009)], 
as well as oscillatory shear [Brader \textit{et al.} (2010)].

Both the full [Brader \textit{et al.} (2008)] and schematic [Brader \textit{et al.} (2009)]
mode-coupling
theories predict an idealized glass 
transition at sufficiently high coupling strength, characterized by an infinitely slow structural 
relaxation time $\tau_{\alpha}$. Ageing dynamics are neglected. 
An important prediction of the approach is that application of any steady strain-rate leads to 
fluidization of the arrested microstructure, with a structural relaxation time determined by the
characteristic 
rate of flow $\tau_{\alpha}\sim\dot\gamma^{-1}$. 

In recent experiments on various soft glassy materials, Ovarlez \textit{et al.} have
indicated that when a dominant, fluidizing shear flow is imposed, then the sample responds as a liquid to an 
additional perturbing shear flow, 
regardless of the spatial direction in which this perturbation is applied. 
These findings imply that once the yield stress has been overcome by the dominant shear flow, 
arrested states of soft matter become simultaneously fluidized in all spatial directions. In
particular, 
the low shear viscosity in a direction orthogonal to the primary flow is determined by the primary
flow rate. 
The rheometer employed in [Ovarlez (2010)] consisted of two parallel discs which enabled the 
simultaneous application of rotational and squeeze shear flow, with independent control over the two 
different shear rates. 
Although this set-up indeed provides a useful way to study superposed shear flows of differing rate, it does not 
provide a mean to test the three-dimensional yield surface, as claimed in [Ovarlez (2010)]. 
A true exploration of the yield surface poses a considerable challenge to experiment and requires a 
parameterization of the velocity gradient tensor which can incorporate the entire family of homogeneous flows, 
including both extension and shear as special cases. 
The superposition of two shear flows is yet another shear flow and does not enable the entire space of 
homogeneous velocity gradients to be explored.

In the present work we will employ the constitutive theory of Brader \textit{et al.} (2009) to
investigate the response of a 
generic colloidal glass to a `mixed' flow described mathematically by the linear superposition of
two independently controllable velocity gradient tensors. 
Numerical results will be presented for the special case in which simple shear is combined with uniaxial
compression. 
Despite the fact that we employ a combination of compression and shear, as opposed to the superposition 
of two shear flows, our theoretical results are broadly consistent with the experimental findings of Ovarlez 
\textit{et al.} (2010) regarding the response of shear fluidized glasses. 
In particular, our calculations reveal clearly the relevant timescales dictating the three dimensional 
response of the system.   
Following this specific application, we proceed to extend our description to treat more general
mixed flows.

The paper will be organized as follows: 
In Sec. \ref{model} we will introduce the deformation measures required to describe flow in 
three dimensions and summarize the schematic model of Brader \textit{et al.} (2009). 
In Sec. \ref{coupled_flows} we will consider the application of our constitutive model to a
specific 
mixed flow, namely a combination of uniaxial compression and simple shear. 
In Sec. \ref{results} we will present numerical results for the flow curves and low shear
viscosity 
for the aforementioned flow combination. 
In Sec. \ref{viscosity} we will perform a perturbation analysis of our constitutive equation
which
enables 
us to address the general problem of superposing a mechanical perturbation onto a dominant flow.  
Finally, in Sec. \ref{discussion} we will discuss the significance of our results and give
concluding 
remarks.


\section{The schematic model}\label{model}


\subsection{Continuum tensors}
Spatially homogeneous deformations are encoded in the spatially translationally
invariant deformation
tensor $\E(t,t')$. 
Any given vector $\rr(t')$ at time $t'$ may be transformed into a new vector $\rr(t)$ at later
time $t$ using the linear relation
\begin{equation}
\rr(t) = \E(t,t')\cdot\rr(t')\,, 
\end{equation}
where $E_{\alpha\beta}=\partial r_{\alpha}(t)/\partial r_{\beta}(t')$
[Brader (2010)]. Calculating the
time derivative of the deformation tensor $\E$ and using the chain rule for derivatives yields
an equation of motion for the deformation tensor
\begin{equation}
\frac{\partial \E(t,t')}{\partial t} = \kap(t)\cdot \E(t,t')\,, 
\label{motionE}
\end{equation}
where $\kap = \nabla\boldsymbol{v}$ is the velocity gradient tensor with components  
$(\nabla\boldsymbol{v})_{\alpha\beta} = \partial \dot{r}_{\alpha}/\partial r_{\beta}$. 
In the present work we will assume incompressibility, which may be expressed by the condition 
${\rm Tr}\,\kap=0$ or, equivalently, ${\rm Tr}\,\E=1$ (volume is conserved).
If the deformation rate is constant in time, then the velocity gradient matrix $\kap$ loses its time
dependence ($\kap(t)\rightarrow \kap$) and the deformation tensor $\E$ becomes a function 
of the time difference alone ($\E(t,t')\rightarrow\E(t-t')$). 
The formal solution of Eq.(\ref{motionE}) for such steady flows is thus given by
\begin{equation}
\E(t)=e^{\kap t}\,.
\label{sol}
\end{equation}
The deformation tensor contains information about both the stretching and rotation of material 
lines (vectors embedded in the material). 
A more useful measure of strain is the Finger tensor $\B(t,t')$, which is defined for steady flows 
by
\begin{equation}
\B(t)=\E(t)\cdot\E^{T}(t)\,.
\label{finger}
\end{equation}
The Finger tensor is invariant with respect to physically irrelevant solid body rotations of the
material 
sample and occurs naturally in many constitutive models (e.g. the Doi-Edwards model of polymer
melts 
[Doi and Edwards (1989)]).


\subsection{Schematic mode-coupling equations}
The schematic model developed in [Brader \textit{et al.} (2009)] expresses the deviatoric stress
tensor
in 
integral form
\begin{equation}
\sig(t)=\int_{-\infty}^{t}\!\!dt'\Big[-\frac{\partial}{\partial t'}\B(t,t')\Big]G(t,t')\,.
\label{constit1}
\end{equation}
An equation of the form (\ref{constit1}) has been derived from first principles 
[Brader \textit{et al.} (2008)], starting from the $N$-particle Smoluchowski equation and applying 
mode-coupling approximations to a formally exact generalized Green-Kubo relation for the stress tensor.  
In [Brader \textit{et al.} (2009)] the theory was simplified to (\ref{constit1}) by assuming spatial 
isotropy of the modulus $G(t,t')$. 
The physical content of Eq.(\ref{constit1}) is that, in order to calculate the stress at the present time,  increments of an appropriate, material objective 
strain measure (the Finger tensor) are integrated over the flow history, each weighted with a 
`fading memory'. 
Approximating $G(t,t')$ by an exponential recovers the well-known Lodge equation [Larson (1988)], which 
is just the integral form of the upper-convected Maxwell model. 
However, Eq.(\ref{constit1}) differs from the simple Lodge equation in that, (i) the modulus $G$ is 
generally not time translationally invariant, due time-dependent variation of the flow in the time interval 
between $t$ and $t'$, (ii) the memory does not decay exponentially to zero, but displays the two-step 
relaxation characteristic of dense colloidal dispersions.

Within the wavevector dependent approach of [Brader \textit{et al.} (2008)] the autocorrelation
function
of stress 
fluctuations is assumed to relax in the same way as the density fluctuations. 
This leads to an approximation for the nonlinear modulus $G$, given 
by a weighted ${\bf k}$-integral over a bilinear function of density correlators at two different 
(but coupled) wavevectors. The schematic model replaces this with the simpler form 
\begin{equation}\label{modulus}
G(t,t')=\nu_{\sigma}\Phi^2(t,t')\,,
\end{equation}
where $\Phi(t,t')$ is a single mode transient density correlator (normalized to 
$\Phi(t,t)=1$) and $\nu_{\sigma}$ is a parameter measuring the strength of stress 
fluctuations.

The dynamics of the single mode density correlator are determined by a nonlinear 
integro-differential equation  
\begin{equation}
\dot{\Phi}(t,t_0) + \Gamma\Bigg\{ \Phi(t,t_0) + \int_{t_0}^{t}dt'
m(t,t',t_0)\dot{\Phi}(t',t_0)\Bigg\} = 0\,,
\label{stdeq}
\end{equation}
where $\Gamma$ is an initial decay rate, the inverse of which sets our basic unit of time. 
The function $m(t,t',t_0)$ is a three-time memory-kernel which depends upon the strain 
accumulated between its time arguments and describes how this competes with the slow structural 
relaxation arising from the colloidal interactions. 
The memory kernel is given by
\begin{equation}
m(t,t',t_0)=h_1(t,t_0)h_2(t,t')\Bigg[\nu_1 \Phi(t,t') + \nu_2 \Phi^2(t,t') \Bigg]\,.
\label{memory}
\end{equation} 
The dependence of the memory upon $\Phi(t,t')$ is taken from the $F_{12}$ model 
developed by G\"otze [G\"otze (2008)]. 
The coupling constants are given by 
$\nu_1=2(\sqrt{2}-1) + \varepsilon/(\sqrt{2}-1)$ and $\nu_2 = 2$, where $\varepsilon$ is a 
parameter expressing the distance to the glass transition.
The system is fluid for $\varepsilon<0$ and in a glassy state for
$\varepsilon>0$.

The $h_i$ entering (\ref{memory}) are decaying functions of the accumulated strain. 
For simplicity we assume $h_1=h_2=h$. 
To allow consideration of any kind of flow (not only shear), the function $h$ is taken 
to depend upon the two invariants $I_1$ and $I_2$ of the Finger tensor
\begin{equation}
h(t,t_0) = \frac{\gamma_{\rm cr}^2}{\gamma_{\rm cr}^2+\Big[\nu I_1(t,t_0) +
(1-\nu)I_2(t,t_0)-3\Big]}\,,
\label{h}
\end{equation}
where a mixing parameter $(0\leq \nu \leq 1)$ and a cross-over strain parameter $(\gamma_{\rm cr})$
have
been introduced [Brader \textit{et al.} (2009)]. The scalars $I_1 =Tr(\B)$ and $I_2 = Tr(\B^{-1})$
are
the trace of the
Finger tensor and its inverse, respectively. In principle, the time evolution of the density 
correlator $\Phi(t,t_0)$ and thus, via Eqs.(\ref{constit1}) and (\ref{modulus}), the stress
tensor, can be calculated by solving Eq.(\ref{stdeq}) numerically for any given 
velocity gradient tensor $\kap$.

The model outlined above contains a set of five independent parameters 
$(v_{\sigma},\Gamma,\varepsilon,\nu,\gamma_{\rm cr})$. 
The least important of these is $\nu$, which determines the relative influence of 
the invariant $I_1$ with respect to $I_2$ in determining the strain induced decay of the memory 
function. 
However, numerical results prove to be extremely insensitive to the value of $\nu$, at least for all 
flows to which the schematic model has so far been applied.  

Trivial scaling of stress and time scales is provided by the parameters $v_{\sigma}$ and 
$\Gamma$.
A statistical mechanical calculation of the dynamics of $N$ colloids (in the absence of 
hydrodynamic interactions) identifies the modulus $G$ as the autocorrelation function of stress 
fluctuations. 
$v_{\sigma}$ therefore determines the initial value of the modulus and, via (\ref{constit1}), 
sets the overall stress scale. 
The reciprocal of the initial decay rate $\Gamma^{-1}$ simply acts as the fundamental timescale. 
For the purpose of our theoretical investigations both $v_{\sigma}$ and $\Gamma$ can, without 
loss of generality, be set equal to unity.  
The theoretical results thus generated can then be fit to experimental data by scaling stress and time 
(or frequency) with alternative values for these two parameters  
[Brader \textit{et al.} (2010)]

The two most important parameters in the model are $\gamma_{\rm cr}$ and $\varepsilon$.
The cross-over strain $\gamma_{\rm cr}$ sets the strain value at which elastic response gives 
way to viscous flow. For example, in experiments considering the shear stress response of dense colloidal 
systems to the onset of steady shear flow, $\gamma_{\rm cr}$ can be identified from the peak of 
the overshoot on the stress-strain curve. 
The parameter $\varepsilon$ characterizes the thermodynamic state point of the system relative to the glass 
transition and serves as proxy for the true thermodynamic parameters of the physical system (volume 
fraction, temperature etc.). 
For example, in a simple system of hard-sphere colloids of volume fraction $\phi$ one can identify 
$\varepsilon\sim(\phi-\phi_{\rm g})/\phi_{\rm g}$, where $\phi_{\rm g}$ is the volume fraction at the glass 
transition. 
For more complicated systems $\varepsilon$ can be regarded as a general coupling parameter which, in the 
absence of flow, yields fluid-like behaviour for $\varepsilon<0$ and amorphous solid-like response for 
$\varepsilon>0$.


\section{Mixed shear and compressional flows}\label{coupled_flows}
With the constitutive relation (\ref{constit1}), we are in a position to determine 
the rheological behaviour of a colloidal glass undergoing any type of homogeneous deformation. 
In [Ovarlez \textit{et al.} (2010)], Ovarlez {\em et al.} considered various soft glassy materials
loaded between two 
parallel discs. 
Each sample was simultaneously sheared by rotating the upper disc about its axis at a given 
angular velocity and squeezed by lowering the height of the upper disc at a given rate. 
By independently varying the rotation and compression rates the stress could be determined 
as a function of one of the rates, for a fixed value of the other. 
In these experiments, the rotation of the upper plate induces a shear flow in the $\hat{{\boldsymbol
\phi}}$ 
direction (in cylindrical coordinates), the rate of which increases linearly with radial distance
from 
the axis of rotation. 
As a consequence of the stick boundary conditions the compression of the sample leads to 
an inhomogeneous shear flow in the $\hat{{\bf r}}$ direction (somewhat akin to a Poiseuille flow)
with 
a maximum shear rate at the boundaries and zero shear rate in the plane equidistant between the two
plates.

\begin{figure}
\begin{center}
\includegraphics[width=0.48\textwidth]{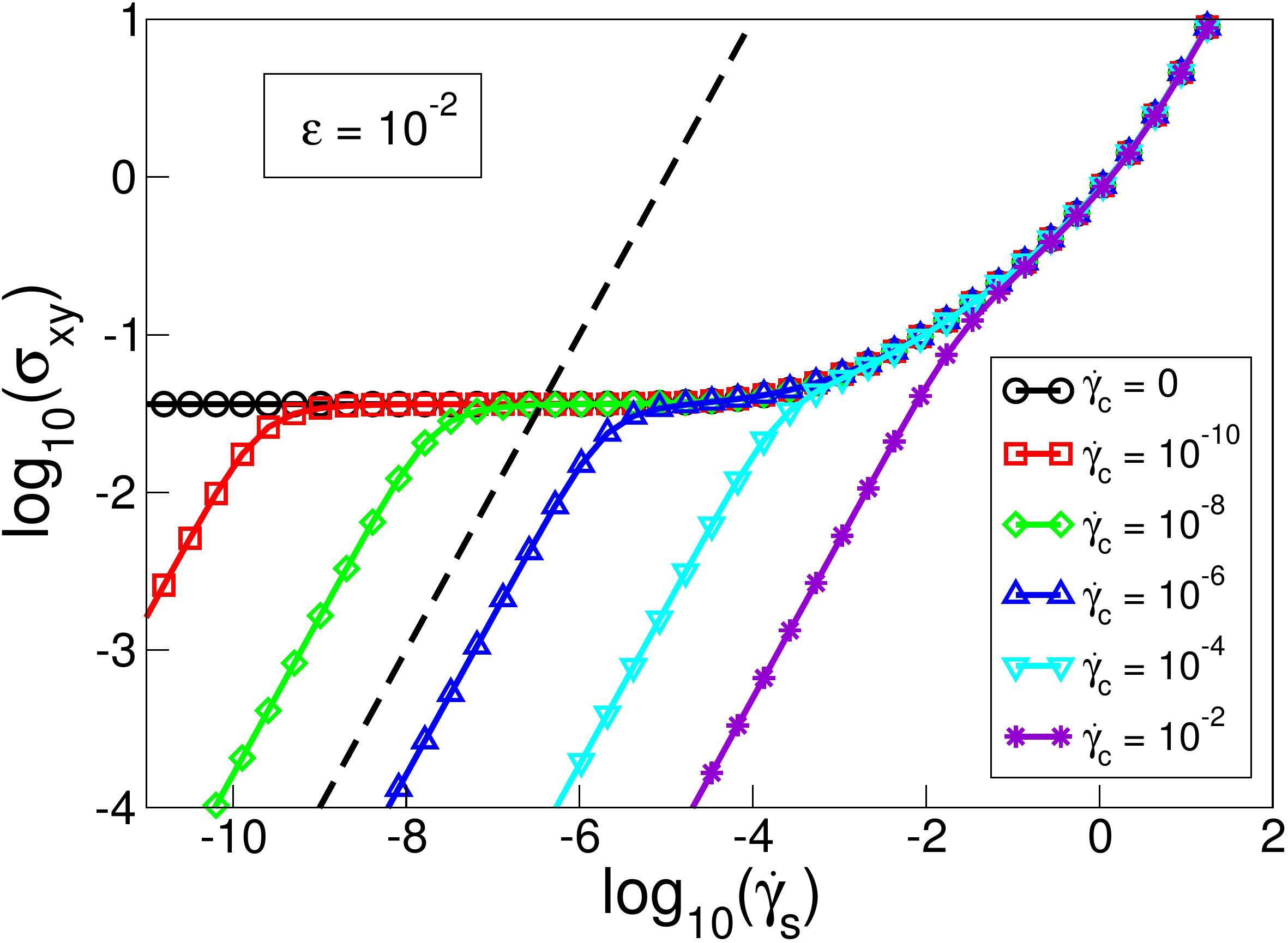}
\caption{The flow curves of a glassy state ($\varepsilon > 0$) for various values of the 
compressional rate $\gamc$. 
The dashed line is a Newtonian viscous law. 
For $\gamc=0$ the $\gams\rightarrow 0$ limit of the flow curve identifies the dynamic yield stress 
[Brader {\em et al.} (2009)]. 
}\label{flow_curves}
\end{center}
\end{figure}

The experiments of Ovarlez \textit{et al.} (2010) were performed in a curvilinear geometry using a
flow
protocol 
which induces an inhomogeneous velocity gradient tensor. 
In principle, spatial variations of the velocity gradient could be treated within the present 
theoretical framework by assuming that the constitutive relations remain valid locally and
enforcing 
the local stress balance appropriate to the geometry of the rheometer under consideration. 
In addition to the increased numerical resources required for such an investigation, the local
application of our 
constitutive equation would represent a further approximation, over and above those already
underlying 
the schematic model. 
The main conceptual point emerging from the experimental studies of Ovarlez \textit{et al.} (2010)
is
that if a
primary flow restores ergodicity and fluidizes the glass, then the response to the secondary flow is
also fluid like. 
Spatial inhomogeneity of one or both flows is merely a complicating factor. 
We thus choose to focus on a more idealized homogeneous flow which is convenient for numerical 
implementation, but nevertheless captures the salient features of the experiment in a minimal way.

The homogeneous flow we choose to implement is a superposition of simple shear and uniaxial 
compressional flow. 
We anticipate that the key physical mechanism at work in fluidized systems under superposed flow 
is the competition between the two imposed relaxation timescales. As the superposition of two shear
flows is itself another shear flow, the experiments of Ovarlez 
\textit{et al.} (2010) leave open the possibility that the observed phenomena could be a  
special feature of shear. 
For this reason we chose to implement the mathematically more general case of superposed extension 
and shear, for which the geometrical coupling of the flows is more involved.

Working in a cartesian coordinate system our flow is specified by
\begin{equation}
\kap = \kap_{s}+ \kap_{c}.
\label{mixed}
\end{equation}
The shear and compressional flows are represented by the following matrices
\begin{equation}
\kap_{s} = \left(
\begin{array}{ccc}
0 & \gams & 0\\
0 & 0 & 0\\
0& 0&0\\
\end{array}
\right)\quad \quad
\kap_{c} = \left(
\begin{array}{ccc}
\gamc/2 & 0 & 0\\
0 & -\gamc & 0\\
0& 0&\gamc/2\\
\end{array}
\right)
\label{comp_shear}
\end{equation}
where $\gams$ and $\gamc$ are the shear and compression rates, respectively. 
Our choice of flow thus differs from those of Ovarlez in two respects, 
(i) both $\gams$ and $\gamc$ are translationally invariant and, 
(ii) we superpose shear with genuine elongation, as opposed to superposing two shear flows.
We consider the flow (\ref{mixed}) as a thought experiment intended to highlight the fundamental
physical 
mechanism of fluidization in a simple and transparent fashion. 
A direct experimental realization of (\ref{mixed}) is not feasible, as this would require a
rheometer 
with stick boundary conditions for generating the shear flow, but slip boundaries for the
compressional 
flow. 
As we will see below, our assumptions do not seem to lead to 
qualitative differences between our theoretical findings and the experimental results and
simplify considerably the theoretical calculations.

Eq.(\ref{sol}) enables calculation of the deformation tensor $\E(t)$ for our mixed flow. 
The non-zero elements are given by 
\begin{eqnarray}
&& E_{xx}=E_{zz}=e^{\gamc t/2}\label{diagonal1}\quad,
\notag\\[0.25cm]
&& E_{yy}=e^{-\gamc t}\label{diagonal2}\quad,
\notag\vspace*{-0.5cm}\\
&& E_{xy}=\frac{2\gams}{3\gamc}e^{-\gamc t}\left( e^{\frac{3\gamc
t}{2}}-1\right)\,.\label{nontrivial}
\end{eqnarray}

\begin{figure}
\begin{center}
\includegraphics[width=0.475\textwidth]{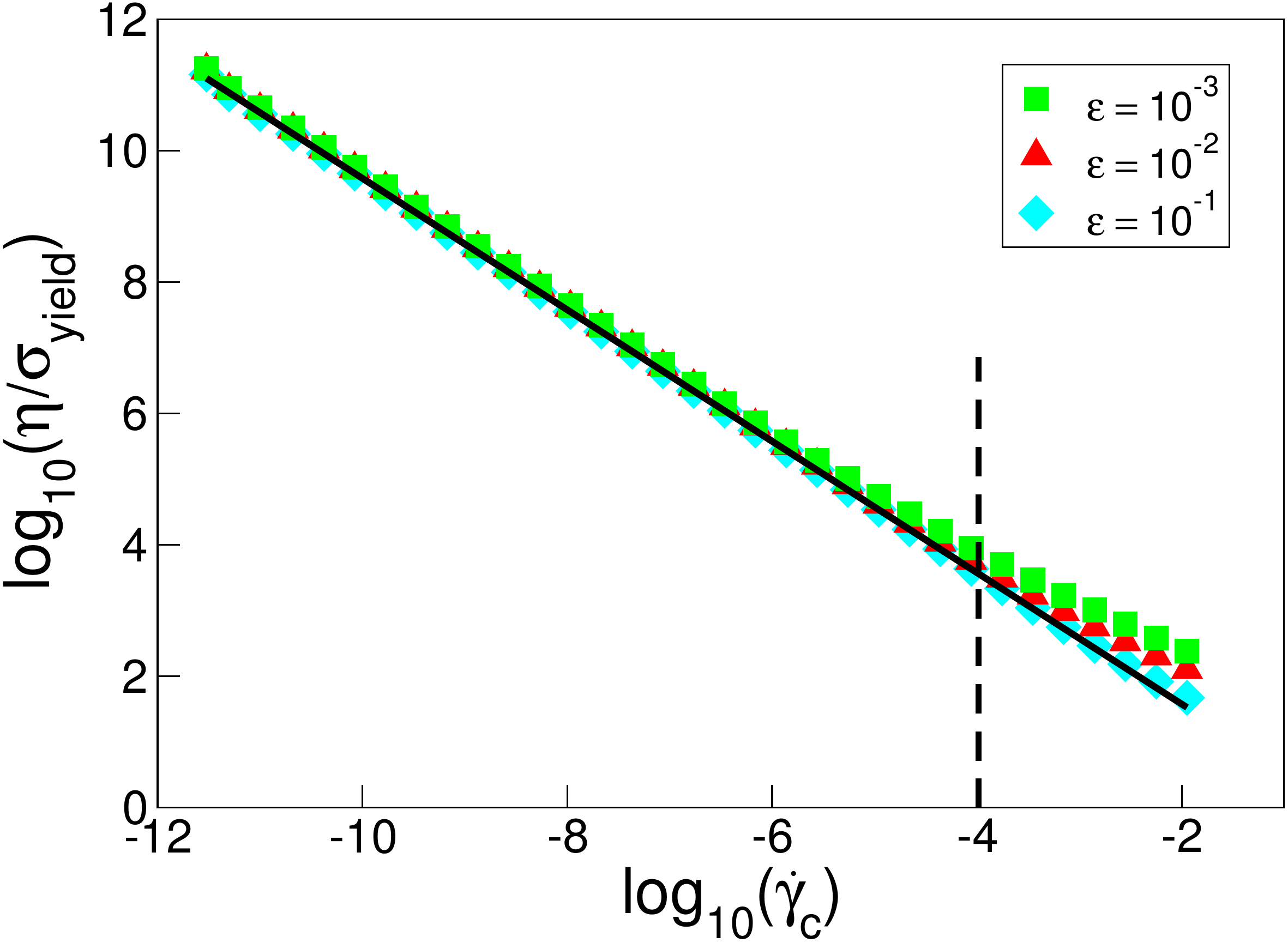}
\caption{The low shear viscosity $\eta$ scaled by
the yield stress $\sigma_{\text{yield}}$ as a function of compression rate $\gamc$ 
(glassy states with $\varepsilon = 10^{-1}, 10^{-2}$ and $10^{-3}$). The continuous line is a
power-law fit to
the numerical data points for $\varepsilon=10^{-1}$ over the range $\gamc=10^{-12}-10^{-4}$ 
and yields an exponent of -1. 
The $\varepsilon$-dependent deviations apparent for $\gamc>10^{-4}$ indicate that short-time 
relaxation processes are becoming relevant. 
}\label{visc_comp}
\end{center}
\end{figure}

\noindent
Employing Eq.(\ref{finger}) yields the Finger tensor
\begin{equation}
\B(t)=\left(
\begin{array}{ccc}
E_{xx}^2+E_{xy}^2 \hspace{0.25cm}& E_{xy}E_{yy} \hspace{0.25cm}& 0\\[0.25cm]
E_{xy}E_{yy} & E_{yy}^2 & 0\\[0.25cm]
0 & 0 & E_{zz}^2
\end{array}
\right)\quad ,
\label{fing}
\end{equation}
with inverse given by
\begin{equation}
\B^{-1}(t)=\left(
\begin{array}{ccc}
\frac{1}{E_{xx}^2} \hspace{0.3cm}& \frac{-E_{xy}}{E_{xx}^2E_{yy}} \hspace{0.25cm}& 0\\[0.3cm]
\frac{-E_{xy}}{E_{xx}^2E_{yy}} & \frac{E_{xx}^2E_{zz}^2 +
E_{xy}^2E_{zz}^2}{E_{xx}^2E_{yy}^2E_{zz}^2} & 0\\[0.3cm]
0 & 0 & \frac{1}{E_{zz}^2}
\end{array}
\right)\quad .
\label{fing-1}
\end{equation}
The invariants required for the memory function prefactors (\ref{h}) are thus 
\begin{eqnarray}
&& I_1(t)=2e^{\gamc t} + e^{-2\gamc t} + E_{xy}^2\label{I1}\quad,\\
&& I_2(t)=2e^{-\gamc t} + e^{2\gamc t} + E_{xy}^2e^{\gamc t}\quad.\label{I2}
\end{eqnarray}
Finally, we need to calculate the time derivative of the Finger tensor $\B(t)$. 
In Sec. \ref{results} we will present results for the shear stress 
$\sigma_{xy}$ as a function of $\gams$, treating $\gamc$ as a parameter. 
Inspection of (\ref{constit1}) shows that we require only the $xy$ component of the 
Finger tensor time derivative
\begin{equation}
\frac{\partial B_{xy}(t)}{\partial t} = \frac{\gams}{3}e^{-2\gamc t}\left( 4-e^{\frac{3\gamc t}{2}}
\right)\quad .
\label{DBxy}
\end{equation}

Substituting (\ref{DBxy}) into (\ref{constit1}) and assuming time translational 
invariance (as appropriate for the steady flows under consideration) we obtain our 
final expression
\begin{equation}
\sigma_{xy} =\! \int_{0}^{\infty}\!\!\!dt\Bigg[\frac{\gams}{3}e^{-2\gamc t}\left( 4-e^{\frac{3\gamc
t}{2}}
\right) \Bigg]\nu_{\sigma}\Phi^2(t).
\label{sigxy}
\end{equation}
The $xy$ component of the shear stress tensor is now completely characterized.
When numerically evaluating the integral in (\ref{sigxy}) we find that truncation at 
$\tau\sim (\gams+\gamc)^{-1}$ provides accurate results. We note that, in an
analogous way, all
other components of the shear stress $\sig(t)$ can be calculated, which is useful if one is
interested for example in the first and second normal stress differences,
$N_1\equiv\sigma_{xx}-\sigma_{yy}$ and $N_2\equiv \sigma_{yy}-\sigma_{zz}$, respectively.


\section{Numerical results}\label{results}

In Fig.\ref{flow_curves} we show flow curves generated from numerical solution of 
Eqs.(\ref{stdeq}-\ref{h}) and (\ref{sigxy}). 
For each curve we set the compressional rate equal to a fixed value, in effect treating $\gamc$ 
as a parameter, and plot the shear stress $\sigma_{xy}$ as a function of $\gams$. 
The model parameters used to generate these data are as follows:
$(\Gamma=1, \nu_{\sigma}=1, \gamma_{\rm cr}=1, \nu=0.5, \varepsilon=10^{-2})$.
For $\gamc=0$ we recover the simple shear flow curve which, for the glassy state under
consideration, 
tends to a dynamic yield stress $\sigma_{xy}\rightarrow\sigma_{\rm yield}=0.03633$ 
in the limit of vanishing shear rate. 
Within the theory, the existence of a dynamic yield stress is a direct consequence of the scaling
of 
the structural relaxation time with shear rate, $\tau_{\alpha}\sim\gams^{-1}$. 
The flow curves calculated at finite $\gamc$ differ qualitatively from that at $\gamc=0$. 
In particular, $\sigma_{\rm yield}$ is a discontinuous function of 
the parameter $\gamc$, such that 
$\sigma_{xy}(\gams\rightarrow0,\gamc=0)\ne \sigma_{xy}(\gams\rightarrow0,\gamc\rightarrow 0^{+})$. 
For finite $\gamc$ values the flow curves present a Newtonian regime for rates $\gams<\gamc$,
followed 
by a shear thinning regime for $\gams>\gamc$. 
The existence of two regimes is quite intuitive: For $\gams<\gamc$ compression is the dominant, 
i.e. fastest, flow and sets the timescale of structural relaxation, whereas for $\gams>\gamc$ the 
shear flow dominates and the flow curve converges to the $\gamc=0$ result.  

\begin{figure}
\vspace{0.5cm}
\begin{center}
\includegraphics[width=0.48\textwidth]{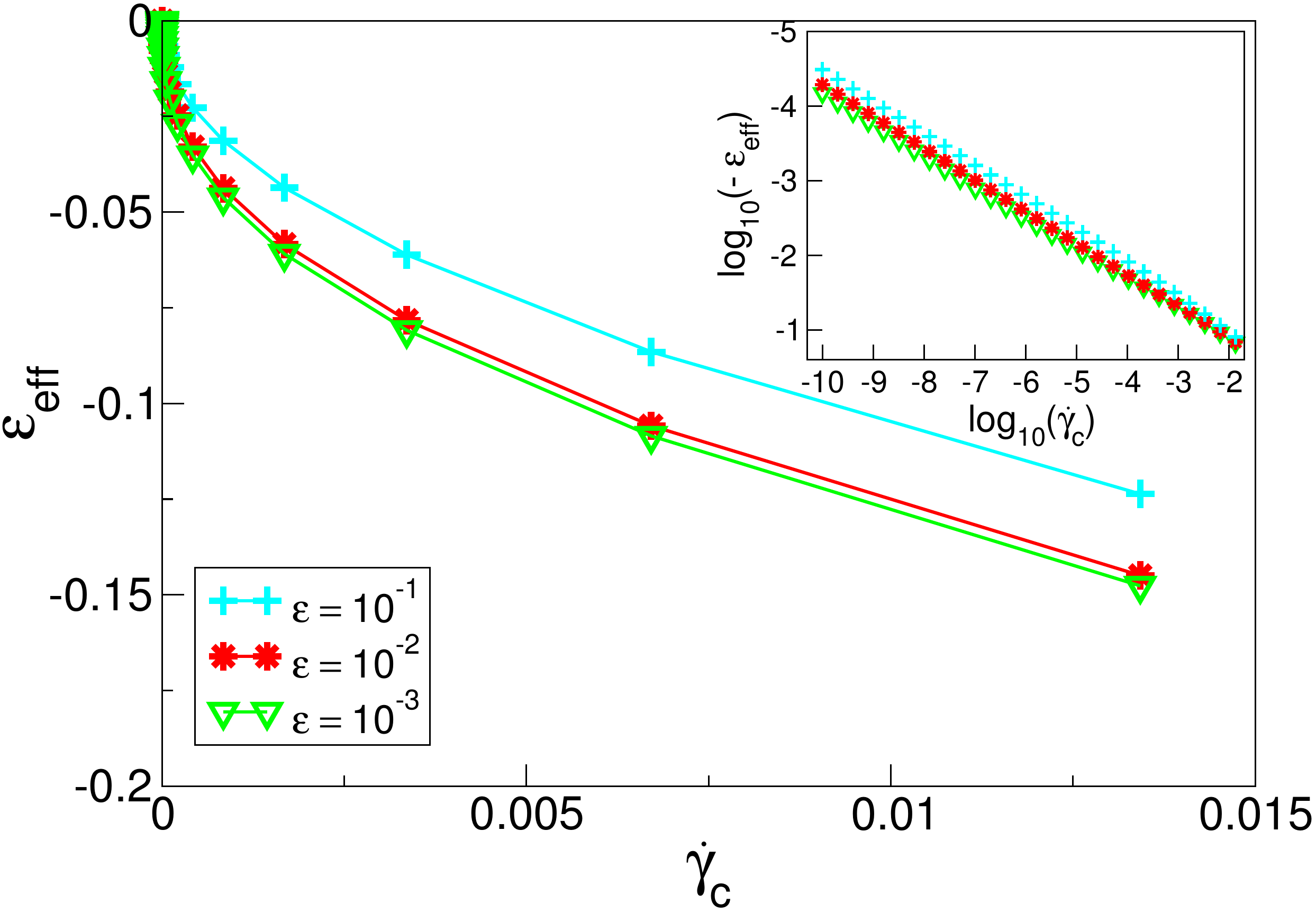}
\caption{Mapping a compression melted glassy state onto an effective fluid state obtained by 
matching the low shear viscosity. 
The effective distance to the glass transition $\eff$ is shown as a function of
the compressional rate $\gamc$, for different values of $\varepsilon$.  
The points are numerical data and the lines provide a guide for the eye.
The inset shows the same data on a logarithmic scale.
}\label{eff_eps}
\end{center}
\end{figure}

The above findings are in good qualitative agreement with the experimental results obtained in 
[Ovarlez \textit{et al.} (2010)] (cf. Fig.3 therein). 
In order to characterize more precisely the flow curves at finite $\gamc$ we show in 
Fig.\ref{visc_comp} the low shear viscosity $\eta=\sigma_{xy}/\gams$, scaled by the yield stress
$\sigma_{\text{yield}}$, 
as a function of $\gamc$ for three different positive values of $\varepsilon$. 
For $\gamc<10^{-5}$ we find very good data collapse onto a master curve. 
For $\gamc>10^{-4}$ clear deviations from universality set in, signifying that the 
compression induced structural relaxation processes are occurring on a timescale within the 
microscopic regime, for which $\eta$ becomes an $\varepsilon$ independent quantity 
(around $\gams=10^{-4}$ for 
the parameter set used in Fig.\ref{flow_curves}). 
Provided that $\gamc<\tau_{\beta}^{-1}$ we find that the numerical data are well 
represented by the power-law scaling 
\begin{equation}
\label{rel_visc_comp}
\eta/\sigma_{\text{yield}}=A\gamc^{\alpha}\,,
\end{equation}
with $\alpha=-1$, in agreement with the experimental findings of Ovarlez \textit{et al.} (2010). 
The constant of proportionality $A$ is independent of $\varepsilon$ (both $\eta$ and 
$\sigma_{\text{yield}}$ vary in the 
same way with this parameter). 
Given the lack of detailed material specificity in the schematic model, we are led to believe that 
$\alpha=-1$ is a universal exponent, independent of both the details of the material under
consideration 
and of the precise nature of the primary and perturbing flows. 
Our findings suggest that any constitutive theory capable of describing a 
three dimensional dynamic yield stress (`yield stress surface' [Brader \textit{et al.} (2009)]) will
inevitably recover 
the scaling (\ref{rel_visc_comp}) with $\alpha=-1$, when applied to tackle mixed flows. 
In particular, we anticipate that the full wavevector dependent mode-coupling constitutive 
equation [Brader \textit{et al.} (2008)] would predict the same scaling behaviour, although this
claim
remains to be 
confirmed by explicit calculations. 
Within mode-coupling-based approaches the value of the scaling exponent $\alpha$ is a natural
consequence 
of the way in which strain enters the memory function (\ref{memory}).

The flow curves presented in Fig.\ref{flow_curves} for various values of $\gamc$ are very
reminiscent 
of the (more familiar) flow curves either measured or calculated under simple shear with $\gamc=0$ 
and $\varepsilon<0$, i.e. states which would remain fluid in the absence of flow (see, e.g.
Fuchs and Cates (2003)). 
This similarity suggests that it may be possible to map, at least approximately, the shear 
response of a steadily compressed, glassy system with $\gamc\ne0$ and $\varepsilon>0$ onto an
uncompressed, fluid 
system, $\gamc=0$, at some effective, negative value of the separation parameter $\eff$. 
One possible way to realize such a mapping is to adjust $\eff$ for a given $\gamc$ to 
obtain equal values for the low shear viscosity of the compressed glass and effective fluid
systems. 
The results of performing this procedure for three values of $\varepsilon$ are shown in 
Fig.\ref{eff_eps}. 
It should be noted that the mapping between $\eff$ and $\gamc$ becomes 
discontinuous at $\gamc=0$ at which point $\eff=\varepsilon>0$. 
The inset of Fig.\ref{eff_eps} shows the same data on a 
logarithmic scale.
In this representation it becomes apparent that the data follow a power law
\begin{equation}
\label{rel_eff}
\eff\sim-\gamc^{\beta}\,.
\end{equation}
Fits to our numerical data yield values for the exponent $0.41<\beta
<0.43$. 

Within the quiescent $F_{12}$ schematic model [G\"otze (2008), G\"otze (1984)], to which the present
theory 
reduces in the absence of flow, it is known that the zero shear viscosity exhibits a power law
divergence 
as $\varepsilon$ approaches the glass transition from below 
\begin{equation}
\label{rel_visc_eps}
\eta\sim (-\varepsilon)^{-\delta}\,,
\end{equation}
where $\delta$ is the same exponent as that describing the divergence of $\tau_{\alpha}$ at the
glass 
transition. Note that the symbol $\delta$ is employed here for this exponent, rather than the
standard choice 
$\gamma$, in order to avoid confusion with the strain.  
When employing the Percus-Yevick approximation to the static structure factor as input, the
wavevector 
dependent mode-coupling theory predicts that for hard-spheres the viscosity exponent takes the
value 
$\delta=2.46$ (identifying $\varepsilon$ as the volume fraction, relative to the transition point) 
[G\"otze and Sj\"ogren (1992)]. 
Within the present schematic model we obtain $\delta=2.3$ (see also footnote \cite{footnote}). 
Given this information about the divergence of $\eta$ in the quiescent system, the power law
relation 
for the mapping (\ref{rel_eff}) is already implicit in the data shown in Fig.\ref{visc_comp}. 
Using the relations (\ref{rel_visc_comp}) and (\ref{rel_visc_eps}) the relation (\ref{rel_eff}) 
can be deduced, where the exponent $\beta$ is given by $\beta = \alpha/(-\delta)=0.43$, which is 
consistent with the results of our numerical fits.


%

\section{Analytic perturbative results}\label{viscosity}

We have so far focused on the special case of mixed shear and compressional flows. 
For any given value of $\gamc<\Gamma$ we have shown that there exists a Newtonian regime in the
stress response to the shear flow, provided $\gams<\gamc$ (see Fig.\ref{flow_curves})
\cite{footnote2}. In this section, we now consider more general situations for which a second slow
flow $\kap_2$ is 
added to a dominant flow $\kap_1$ (while keeping the requirements of
incompressibility and homogeneity). In the present context a sufficient condition
for the second flow to be considered `slow' is that $\gamtwo\ll\gamone$, where the characteristic
shear rates are now identified as $\dot{\gamma}_i=\sqrt{\kap_i\!:\!\kap^T_i}$ for $i=1,2$ (where
${\bf A}\!:\!{\bf
B}\equiv\sum_{ij}A_{ij}B_{ji}$). In Subsections \ref{newt}-\ref{experiment}, we provide
perturbative constitutive
equations for three different cases. In the first of these cases, we consider $\kap_1$ and $\kap_2$
as steady flows (without any other restriction), derivate the corresponding pertubative constitutive
equation and finally apply this latter to our coupled compressional and shear flows, in order to
theoretically account for the Newtonian viscous response to $\kap_2$ discussed in 
Sec. \ref{results}, and to finally make the connection with the phenomenological constitutive
equation
obtained by Ovarlez \emph{et al.}. In the second case, we still consider steady flows, but this time
with the additional requirement of `commutating' flows, i.e.
$[\kap_1,\kap_2]\equiv\kap_1\cdot\kap_2-\kap_2\cdot\kap_1=0$, whereas in the third case $\kap_2$ is
time-dependent and the requirement of commutating flows is maintained. We also illustrate these
last two cases with instructive examples.



\subsection{Newtonian viscous response}\label{newt}

\subsubsection{Perturbation expansion}

The first point to note is that the isotropic modulus $G(t)$ decays on the timescale 
$\gamone^{-1}$; the slower secondary flow has no influence on the structural relaxation.
We henceforth make this fact explicit in the notation for the modulus by writing 
$G(t;\gamone,\gamtwo)\equiv G(t;\gamone)\equiv G_1(t)$.  
For steady flows the constitutive equation (\ref{constit1}) may thus be simplified to
\begin{eqnarray}
\sig=\int_0^{\infty}\!\!dt\left(
\frac{\partial}{\partial t}\B(t)
\right)G_1(t).
\label{visc3}
\end{eqnarray} 
Although the modulus is essentially independent of $\gamtwo$, the Finger tensor depends 
nonlinearly upon both $\kap_1$ and $\kap_2$. 
In order to address the case $\gamtwo\ll\gamone$ we expand the Finger 
tensor (\ref{finger}) to first order in $\kap_2$. 
For the mixed flow under consideration $\B(t)$ is given by 
\begin{eqnarray}
\B(t)=
e^{
(\kap_1+\kap_2)t
}
e^{
(\kap^T_1+\kap^T_2)t
}.
\label{finger_mixed}
\end{eqnarray}
The desired partial linearization of (\ref{finger_mixed}) with respect to $\kap_2$ is complicated by
the 
fact that the two velocity gradient tensors do not necessarily commute.  

In order to proceed we consider the following Taylor expansion
\begin{eqnarray}
e^{\xh+\alpha \yh} = e^{\xh} + \alpha\left[
\frac{d}{d\alpha}e^{\xh+\alpha \yh}
\right]_{\alpha=0}
\!+\;\mathcal{O}(\alpha^2), 
\label{expansion}
\end{eqnarray}
where $\xh$ and $\yh$ are arbitrary operators independent of the scalar coupling parameter
$\alpha$. 
The derivative may be obtained using the Feynman identity [Feynman (1951)]
\begin{eqnarray}
\left[
\frac{d}{d\alpha}e^{\xh+\alpha \yh}
\right]_{\alpha=0}
=\int_0^1 \!d\lambda\, e^{\xh (1-\lambda)}\yh\,e^{\xh \lambda}.
\label{identity}
\end{eqnarray}
Applying (\ref{expansion}) and (\ref{identity}) to (\ref{finger_mixed}) we obtain the leading order 
result 
\begin{eqnarray}
\B(t)=\B_1(t) + \int_0^t \!\!ds\,\left( \U(t,s) + \U^T(t,s) \right),
\label{visc1}
\end{eqnarray}
where we define the following tensors
\begin{eqnarray}
\B_1(t)&\equiv&\E_1(t)\!\cdot\!\E^T_1(t)
\\
\U(t,s) &\equiv& \E_1(t)\!\cdot\!\E_1(-s)\!\cdot\!\kap_2\!\cdot\!\E_1(s)\!\cdot\!\E^T_1(t),
\label{visc2}
\end{eqnarray}
where $\E_1(t)=\exp(\kap_1 t)$.
Eq.(\ref{visc1}) is linear in $\kap_2$ but retains all orders of the dominant flow $\kap_1$. 
Substitution of (\ref{visc1}) into (\ref{visc3}) thus yields a stress tensor
consisting of two 
terms,
\begin{eqnarray}
\sig &=& \int_{0}^{\infty}\!\!\!\!dt\,\left(\frac{\partial}{\partial t}\B_1(t)\right)G_1(t)\notag\\
&\phantom{=}&+\int_{0}^{\infty}\!\!\!\!dt \frac{\partial}{\partial t}\left[\int_0^t \!\!ds\,\left(
\U(t,s)+ \U^T(t,s) \right)\right] G_1(t)\notag\\
&\equiv&\sig_1 +\delta\sig\label{perturb_1}\quad,
\end{eqnarray}
where $\sig_1$ is the stress arising purely from the dominant flow 
and $\delta\sig$ is the additional contribution from the slow perturbation.

The perturbative term $\delta\sig$ in (\ref{perturb_1}) is a tensor whose elements
depend 
upon time as $[\cdot]_{ij}\sim t^{n_{ij}}$, where $n_{ij}\ge 1$ is an integer. 
Within the schematic model the relaxation time determining the decay of $G_1(t)$ is given by 
$\tau_{\alpha}=\gamma_{\rm cr}/\gamone$, where $\gamma_{\rm cr}$ is the cross-over strain parameter
entering 
(\ref{h}). 
This decay serves to cut off the integral in (\ref{perturb_1}) at the upper limit
$t\sim\tau_{\alpha}$, 
with the consequence that the numerically largest elements of $[\cdot]_{ij}$ arising from terms
with 
$n_{ij}=1$ are $\mathcal{O}(\eta_1\gamma_{\rm cr})$.
In a (repulsive) colloidal glass any given colloid is trapped within a cage of nearest neighbours. 
The cross-over strain parameter $\gamma_{\rm cr}$ is related to the strain at which the cages begin
to be broken by the external flow. Typical values for this dimensionless parameter from simulation
or experiment are $\gamma_{\rm cr}\approx 0.1$ [Zausch \textit{et al.} (2008)].

We now apply the perturbative formula (\ref{perturb_1}) to the coupled
compressional and shear flows expressed by the matrices (\ref{comp_shear}), with $\kap_1=\kap_c$ and
$\kap_2=\kap_s$. Since no complete
analytic expression is known for the density correlator $\Phi$, we approximate the modulus
$G_1(t)$ by an exponentially decaying function 
$G_1(t)\approx G_{\infty}\exp(-\gamone t/\gamma_{\rm cr})$, 
where $G_{\infty}$ is a constant. Under this approximation, Eq. (\ref{perturb_1}) becomes
\begin{eqnarray}
\sig &=& \eta_1\gamc
\left(
\begin{array}{ccc}
\frac{1}{1-\gamma_{cr}} & 0 & 0\\
0 & \frac{-2}{1+2\gamma_{cr}} & 0\\
0 & 0 & \frac{1}{1-\gamma_{cr}}
\end{array}
\right)\notag\\
&\phantom{=}& + 2\frac{2\eta_1}{(2+5\gamma_{cr}+2\gamma_{cr}^2)}\D_2\label{newtonian_term}\quad,
\label{visc_reduction}
\end{eqnarray}
where $\eta_1\equiv\int_{0}^{\infty}dt \,\,G_1(t)$ is the rate dependent shear viscosity of the
primary flow alone and $\D_2\equiv(\kap_2+\kap^{T}_2)/2$ is the
symmetric part of the velocity gradient matrix $\kap_2$. The second term in
(\ref{newtonian_term}) is nothing but the expression of a Newtonian-type viscous response to the
secondary flow $\kap_2$ (shear flow) with a viscosity mainly determined by the strain rate 
of the dominant flow $\gamone$ (compressional rate) through $\eta_1$. This is in agreement
with what we numerically showed in Sec. \ref{results}.


\subsubsection{Empirical constitutive equation}

In [Ovarlez \textit{et al.} (2010)], Ovarlez {\em et al.} proposed an empirical constitutive
equation
to account for the 
viscous stress measured in a number of fluidized glassy systems. 
In the notation of the present work the proposed constitutive relation is 
\begin{eqnarray}
\sig=2\left[
\frac{\sigma_{\rm yield}+kd^n}{d}
\right]\D,
\label{ovarlez_eq}
\end{eqnarray}
where $k$ and $n$ are scalar parameters and $d\equiv \sqrt{2 \D\!:\!\D}$ is an invariant of the
symmetric 
part of the total velocity gradient $\D=(\kap+\kap^T)/2$. 
The isotropic viscosity appearing in square parentheses in (\ref{ovarlez_eq}) is obtained from a
straightforward 
generalization of the familiar scalar Hershel-Bulkley law for the shear stress, 
$\sigma_{\rm sh}=({\sigma_{\rm yield}}+k\dot\gamma^n)$.  

If we neglect the cross-over strain parameter $\gamma_{cr}$ (whose value is already
small, $\gamma_{cr}\approx0.1$), then the second term in our perturbative constitutive equation
(\ref{newtonian_term}) becomes $\delta\sig=2\eta_1\D_2$, which is entirely consistent with the
implicit second term in the empirical relation (\ref{ovarlez_eq}).
Indeed, for $\gamone\gg\gamtwo$ the generalized Hershel-Bulkley effective viscosity $(\sigma_{\rm
yield}+kd^n)/d$ is 
dominated by the fastest flow and is effectively independent of $\gamtwo$. 
We can thus make the following correspondence between viscosities appearing in the schematic
(\ref{newtonian_term}) 
and empirical (\ref{ovarlez_eq}) constitutive equations 
\begin{eqnarray}
\eta_1\equiv\int_0^{\infty}\!\!dt\, G_1(t)\;
\Longleftrightarrow
\;\frac{\sigma_{\rm yield}+kd^n}{d}.
\end{eqnarray}
The linear dependence of $\D$ on the velocity gradient tensor thus enables (\ref{ovarlez_eq}) to be 
rewritten as 
\begin{eqnarray}
\sig=\sig_1 + 2\eta_1\D_2,
\label{ovarlez_eq2}
\end{eqnarray}
thus making explicit the connection between (\ref{newtonian_term}) and (\ref{ovarlez_eq}).


\subsection{Anisotropic viscosity}\label{anisvis}

We still consider steady flows $\kap_1$ (dominant flow) and $\kap_2$ (secondary
flow), but now with the restriction
of commutating flows, i.e. $\left[\kap_1,\kap_2\right]=0$. 
Such flows have the property that the total deformation tensor 
can be formed from the product of the individual deformations, 
$\E(t)=\E_1(t)\E_2(t)$.
As we will see, this restriction allows
for more tractable pertubative constitutive equations.

With $\left[\kap_1,\kap_2\right]=0$, the expression (\ref{visc1}) then reduces to
\begin{eqnarray}
\B(t)=\B_1(t)\, + \;2t\,\E_1(t)\!\cdot\!\D_2\!\cdot\!\E_1^T(t)\quad.
\label{visc4}
\end{eqnarray}
Substitution of (\ref{visc4}) into (\ref{visc3}) yields the following form for the 
stress tensor 
\begin{eqnarray}
\label{perturb_2}
\sig &=& \sig_1 +  2\eta_1\D_2 \label{stress_expansion}\\
&&+\,
2\!\int_0^{\infty}\!\!\!\!dt\,
G_1(t)\left[
\frac{\partial}{\partial t}
\left(
t\,\E_1(t)\!\cdot\!\D_2\!\cdot\!\E_1^T(t)
\right)
\!-\D_2
\right].
\notag
\end{eqnarray}
The anisotropic third term in (\ref{perturb_2}) is the result of a nonlinear
operator acting on the perturbing velocity gradient
$\D_2$ and 
incorporates information about the symmetry imposed on the system by the dominant 
fluidizing flow. 

For incompressible isotropic fluids in the Newtonian regime the viscosity in any given flow can be
determined 
from the shear viscosity via Trouton's rules (e.g. $\eta_{\rm el}=3\eta_{\rm sh}$, where $\eta_{\rm
el}$ is 
the elongational viscosity in uniaxial extension). 
Trouton's rules no longer hold in the present case, due to the presence of the third term in 
($\ref{stress_expansion}$).

In order to explicitly demonstrate the relative magnitude of the anisotropy, we consider the special
case of  perpendicular 
shear flow $(\kap_1)_{ij}=\gamone\delta_{ix}\delta_{jy}$,
$(\kap_2)_{ij}=\gamtwo\delta_{iz}\delta_{jy}$ and 
again approximate the modulus by an exponentially decaying function 
$G_1(t)\approx G_{\infty}\exp(-\gamone t/\gamma_{\rm cr})$.
Under this simplifying assumption Eq.(\ref{stress_expansion}) becomes
\begin{eqnarray}
\sig = \sig_1 + 2\eta_1\gamtwo\!
\left(
\begin{array}{ccc}
0 & 0 & 0\\
0 & 0 & 1/2\\
0 & 1/2 & 0\\
\end{array}
\right)
\!+
2\eta_1\gamtwo\!
\left(
\begin{array}{ccc}
0 & 0 & \gamma_{\rm cr}\\
0 & 0 & 0\\
\gamma_{\rm cr} & 0 & 0\\
\end{array}
\right)\!.
\notag\\
\label{testcase}
\end{eqnarray}
The small off-diagonal elements which appear in the third term of (\ref{testcase}) are generated by
the coupling between primary and perturbing flows and may be viewed as a 
correction, at around the $10$\% level, to the dominant isotropic viscosity $\eta_1$.
The appearance of these additional contributions to the viscous stress can be attributed to the 
normal stress differences generated by the primary flow. Constitutive theories with vanishing normal
stress differences will always predict an isotropic viscous response to perturbing flows. 

It is interesting to note that the stress tensor (\ref{stress_expansion}) 
may be formally expressed in terms of an anisotropic viscosity 
\begin{eqnarray}
(\sig)_{ij} = (\sig_1)_{ij} \;+\; 2(\etab)_{ijkl}(\D_2)_{kl}, 
\label{fourthrank1}
\end{eqnarray}
where the fourth rank tensor $\etab$ with components $(\etab)_{ijkl}$ is given in terms of the
shear 
modulus and the deformation gradient of the dominant flow 
\begin{eqnarray}
(\etab)_{ijkl} =  
2\!\int_0^{\infty}\!\!\!\!dt\,
G_1(t)\left[
\frac{\partial}{\partial t}
t\,(\E_1(t))_{ik}(\E_1(t))_{jl}
\right].
\label{fourthrank2}
\end{eqnarray}
If the dominant flow is switched off, then $\E_1(t)=\one$ and (\ref{fourthrank2}) reduces to the
familiar 
isotropic viscosity $\eta_{ijkl}=2\eta_0\delta_{ik}\delta_{jl}$, where $\eta_0$ is the zero shear 
viscosity (infinite for glassy states with $\varepsilon>0$). 
Given that the dominant flow fixes the anisotropy of the system it is not surprising 
that the viscosity experienced by the perturbing flow is a tensorial quantity.

Finally, we note that the presence of anisotropy prevents a general three dimensional mapping of a
flow fluidized 
glass onto an effective fluid state with $\varepsilon<0$, as performed for the special case of
mixed 
compressional and shear flow in Sec. \ref{results}.

\begin{figure}
\begin{center}
\includegraphics[width=0.475\textwidth]{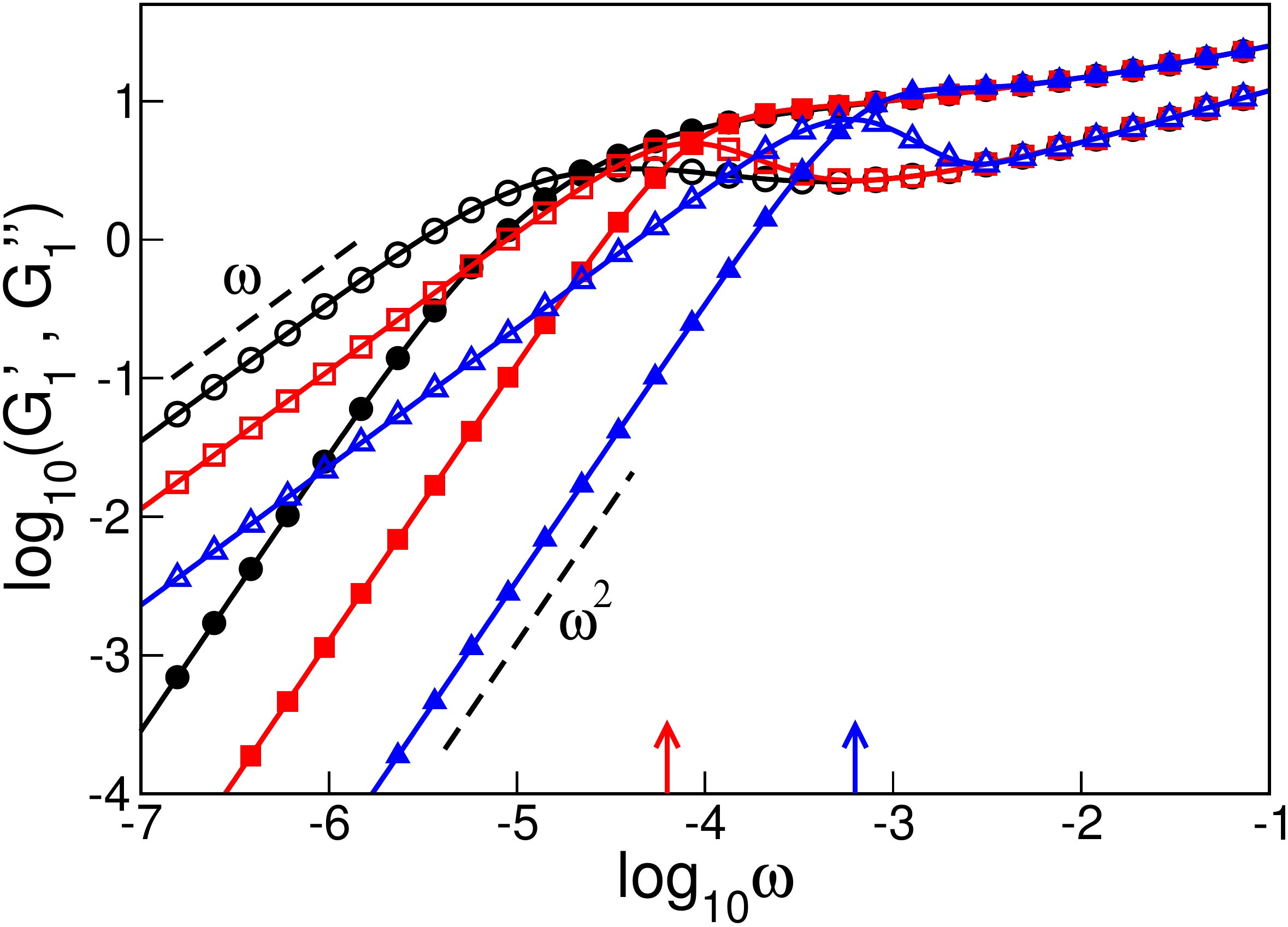}
\caption{The othogonal superposition storage ($G_1'$, filled symbols) and loss ($G_1''$, open
symbols) 
moduli as a function of frequency for a fluid state.  
The (black) circles show the standard linear response moduli calculated at a vanishing steady shear 
rate $\gamone=0$. 
The (red) squares and (blue) triangles show the moduli calculated at finite steady shear rates, 
$\gamone=10^{-5}$ and $10^{-4}$, respectively. The arrows indicate the characteristic frequency 
$2\pi\gamone$  below which the structural relaxation is dominated by the steady shear (which 
coincides approximately with the crossing point of the moduli).  
Parameter set $(\Gamma=1, \nu_{\sigma}=100, \gamma_{\rm cr}=1, \nu=1, \varepsilon=-5\times10^{-3})$.}
\label{moduli}
\end{center}
\end{figure}

\subsection{Superposition spectroscopy}\label{experiment}
A special case of mixed flow which has received some attention in the rheological literature 
is small amplitude oscillation superposed onto steady shear. 
Largely due to constraints imposed by the available apparatus, 
the majority of the experimental works have involved parallel shearing flows 
using either cone-plate 
[Booij (1966), Osaka \textit{et al.} (1965)] or Couette [Vermant \textit{et al.} (1998)]
rheometers. The measured viscoelastic parallel superposition moduli depend upon both the
microstructure under steady shear and its evolution with changes in shear rate. 

A more informative, albeit harder to realize, mixed flow consists of oscillatory shear superposed
perpendicular to the main flow 
direction. 
The orthogonality of the flows makes possible a mechanical spectroscopy of flowing systems which
can 
probe flow induced changes in the microstructure and which may be used as a useful test of
constitutive 
equations [De Cleyn and Mewis (1987), Kwon and Leonov (1993), Leonov \textit{et al.} (1976),
Tanner and Simmons (1967)]. 
For details on the experimental realization of such orthogonal flows we refer the reader to Vermant
\textit{et al.} (1997).

As a specific example of orthogonal oscillation we consider the mixed flow 
\begin{eqnarray}
(\kap_1)_{ij}&=&\gamone\delta_{ix}\delta_{jy}\label{sh}
\\
(\kap_2)_{ij}&=&\gamma_0\,\omega\cos(\omega t)\delta_{iz}\delta_{jy}\label{osc},
\end{eqnarray}
where $\omega$ is the angular frequency and $\gamma_0$ is the amplitude of the oscillatory 
strain (assumed to be small). The time-dependence of the perturbing flow $\kap_2$
requires us to use a time-ordered exponential to express the corresponding deformation
tensor, namely
\begin{equation}
\E_2(t,t') = e_{+}^{\int_{t'}^{t}ds\kap_2(s)}\quad,
\end{equation}
where the exponential is defined according to
\begin{equation}
e_{+}^{\int_{t'}^{t}\!ds A(s)}= \boldsymbol{1} + \int_{t'}^{t}\!\!\!ds A(s) +
\int_{t'}^{t}ds_1\int_{t'}^{s_1}\!\!\!ds_2\,\,A(s_1)A(s_2) + \ldots
\end{equation}
We also require the time-dependent expression for the stress tensor given by (\ref{constit1}). After
linearization with respect to $\kap_2$ and making use of
$\left[\kap_1,\kap_2\right]=0$, we obtain a formula for the stress tensor analogous to
(\ref{stress_expansion}),
\begin{eqnarray}
\sig(t) &&= \sig_1 + \int_{-\infty}^{t}\!\!\!\!\!\!dt'G_1(t,t';\gamone)\notag\\
&&\times\left\{-\frac{\partial}{\partial t'}\left[
\E_1(t-t')\!\cdot\!\left(\int_{t'}^{t}\!\!\!\!ds
\,\,2\,\,\D_2(s)\right)\!\cdot\!\E_1^T(t-t')\right]\right\}.\notag\\
\label{stress_expansion_time}
\end{eqnarray}
Substituting (\ref{sh}) and (\ref{osc}) into
(\ref{stress_expansion_time}) and making use of standard trigonometric addition formulas yields 
\begin{eqnarray}
\sigma_{zy}(t)\!=\!\gamma_0 G'_1(\omega;\gamone)\sin(\omega t) 
+ 
\gamma_0 G''_1(\omega;\gamone)\cos(\omega t),
\notag\\ 
\label{osc_stress}
\end{eqnarray}
where the orthogonal superposition moduli  are given by 
\begin{eqnarray}
G'_1(\omega;\gamone)&=&\omega\!\int_0^{\infty}\!\!dt' \sin(\omega t')\,G_1(t';\gamone), 
\label{gp}
\\
G''_1(\omega;\gamone)&=&\omega\!\int_0^{\infty}\!\!dt' \cos(\omega t')\,G_1(t';\gamone). 
\label{gpp}
\end{eqnarray}
In Eqs.(\ref{osc_stress}-\ref{gpp}) we have made explicit the dependence of the moduli upon the
steady 
shear rate $\gamone$. 
The application of oscillations perpendicular to the flow thus enable the modulus under steady
shear 
to be investigated and provide information about the shear induced relaxation of stress
fluctuations. 
We note that identical moduli (\ref{gp}) and (\ref{gpp}) would be obtained had we chosen the 
alternative perturbing flow $(\kap_2)_{ij}=\gamma_0\,\omega\cos(\omega t)\delta_{ix}\delta_{jz}$ 
and determined the stress component $\sigma_{xz}$.

In Fig.\ref{moduli} we show the orthogonal superposition moduli as a function of frequency for
three 
different values of the steady shear rate $\gamone$. 
For $\gamone=0$ we recover the standard linear response moduli, for which the viscous loss dominates
the 
elastic storage for frequencies less than $2\pi/\tau_{\alpha}$. 
For the fluid state considered $\tau_{\alpha}$ is finite (\textcolor{red}{$\varepsilon<0$}). 
As the steady shear rate is increased, relaxation processes with rates less than $\gamone$ are
suppressed 
and the point at which the storage and loss moduli cross moves to higher frequency. 
These findings are consistent with the experimental results of both Booij (1966) and Vermant
\textit{et
al.} (1998).
The underlying physics here is essentially the same as that leading to the shift of the Newtonian
regime 
shown in Fig.\ref{flow_curves}. 
At low frequencies the orthogonal superposition moduli retain the same frequency scaling as in the
familiar 
unsheared situation, namely $G'_1\sim\omega^2$ and $G''_1\sim\omega$. 
We note also that the Kramers-Kronig relations remain valid for finite values of $\gamone$. 

In addition to the speeding up of structural relaxation induced by the steady shear, the loss
modulus also 
displays a more pronounced $\alpha$-peak compared to the unsheared function. 
This feature is related to the functional form of the $\alpha$-decay of the transient density
correlator. 
In the absence of shear $\Phi(t)$ decays as a stretched exponential, whereas under shear the final
decay is 
closer to pure exponential. 
However, it is likely that more accurate (i.e. beyond schematic) orthogonal moduli, obtained either
from 
experiment or more detailed microscopic calculations/simulations, would differ qualitatively in the
region 
of the $\alpha$-peak. 
There is accumulating evidence [Zausch \textit{et al.} (2008)] that $G_1(t;\gamone)$ becomes
negative at
long times and, as 
this feature is not captured by the simple schematic model employed here, differences in the Fourier
transformed 
quantity around $\omega=2\pi\gamone$ may be anticipated. 
The negative tail of  $G_1(t;\gamone)$ is related to the existence of a maximum in the shear stress
as a function 
of time following the onset of steady shear [Zausch \textit{et al.} (2008)]. 
The `stress overshoot' is present in the full wavevector dependent mode-coupling equations [Brader
\textit{et al.} (2008)] but gets 
lost in simplifying the theory to the schematic level.

\section{Discussion and conclusions}\label{discussion}
In this paper we have demonstrated that the MCT-based schematic model of Brader \textit{et al.}
(2009)
can qualitatively 
account for the 
experimental results on three dimensional flow of soft glassy materials reported in [Ovarlez
\textit{et
al.} (2010)]. 
In particular, the competition of timescales which arises from applying flows of differing rate  
appears to be correctly incorporated into the model. 
The main outcome of our analysis is that the viscous response to a perturbing secondary flow is 
dominated by the primary flow rate. Although subtle anisotropic corrections to this 
picture do emerge from our equations, it remains to be seen whether these have significant 
consequences for experiments in any particular rheometer geometry. 

A key feature of the mode-coupling constitutive theory is that it captures the transition from an ergodic fluid to an 
arrested glass as a function of the coupling strength. 
The present study demonstrates that the theory qualitatively accounts for experimental data on 
mechanically fluidized glassy systems in three dimensional situations. 
What remains to be established is whether the experimental yield surface of a colloidal glass agrees with 
the (almost) von 
Mises form [Hill (1971)] predicted by the schematic model [Brader {\em et al.} (2009)]. 
A true measurement of the yield surface would require a rheometer which enables parameterization of 
the entire family of velocity gradients, incorporating both uniaxial and planar extensional flows. 
This has not yet been achieved. 
Given the very different mechanisms underlying plastic flow in colloidal glasses and metals (for which the 
von Mises yield surface was originally proposed) direct measurements of the yield surface could 
prove very informative.


The good qualitative agreement of our theory with the experimental results of Ovarlez \textit{et
al.}
(2010) on fluidized glasses is perhaps all the 
more surprising when recalling that the theory is constructed specifically for dispersions of 
spherical colloidal particles (without hydrodynamic interactions), whereas the experiments 
were performed on large aspect ratio Bentonite clay, a Carbopol gel and an emulsion. 
The consistent phenomenology presented by these disparate systems would seem to indicate that the 
sufficient elements required for a successful theory are (i) a well-founded geometrical structure 
(in the sense that its tensorial structure is appropriate), 
(ii) correct incorporation of flow induced relaxation rates. 
The specific nature of the interparticle interactions does not seem to be of particular importance, although 
we note that certain interaction potentials may be more susceptible to inhomogeneous flow 
(e.g. shear banding instabilities) than others. 
The presence of a spatially varying velocity gradient tensor $\kap$ would conflict with the assumption of 
translational invariance underlying our constitutive equation.

For the case of superposed compression and shear flow we have found that the viscosity felt by the perturbing 
shear flow is given by $2\eta_1/((2+5\gamma_{cr}+2\gamma_{cr}^2))$, which for the typical value 
$\gamma_{cr}\!=\!0.1$ is around $20$\% less than the primary viscosity $\eta_1$. 
In [Ovarlez \textit{et al.} (2010)] the sedimentation velocity of a sphere falling in the vorticity 
direction of a shear fluidized glass was observed to be a factor of $1.4$ larger than one would expect 
from a sphere falling through a fluid of viscosity $\eta_1$. 
Ovarlez \textit{et al.} attributed this to hydrodynamic interactions between sedimenting particles. 
Although the flow around a falling sphere in shear flow is more complicated than the flows considered 
in the present work, it is nevertheless tempting to speculate that enhanced sedimentation velocity 
could be connected to a reduced effective viscosity, as occurs in (\ref{visc_reduction}), arising from 
a nontrivial coupling of the superposed flows. 
We leave a detailed application of our constitutive equation to the problem of sedimentation under
shear 
to future work.

An aspect of the present work which may warrant further investigation is the possible analogy
between 
systems with isotropic interparticle interactions, upon which anisotropy is imposed by external mechanical 
force fields, and intrinsically anisotropic materials such as liquid crystals
[de Gennes and Prost (1993), Larson (1988)].
The theory of  anisotropic fluids has a long history, beginning essentially with the work of Oseen in 
the 1930s [Oseen (1933)] and developed through the work of Ericksen (1959) and Leslie (1968). 
In all of these theoretical developments the anisotropy of the viscous response originates from the 
underlying anisotropy of the constituents; usually oriented polymers or rod-like particles with 
liquid crystalline order. 
Take the nematic phase as an example. 
Within a continuum mechanics description the orientational order 
is characterized by the director ${\bf n}$. 
In certain systems the director may be held fixed by the application of a suitably strong external field, 
in others it interacts with the imposed flow in a more complicated way. 
Either way, the presence of a preferred direction in the sample gives rise to an anisotropic viscous 
response (characterized e.g. by the five scalar Leslie viscosity coefficients). 
In the present case the anisotropy of the perturbation response is determined by the geometry of the primary 
flow. 
It may thus be anticipated that the eigenvectors of the primary deformation may play a role in the present 
theory analogous to that of ${\bf n}$ in the dynamics of nematics. 
   
Throughout the present work we have focussed on the response of a glass which has already been fluidized 
by a primary flow of constant rate. 
However, within the same formalism we can also consider the predictions of our constitutive equation for 
the elastic response of a colloidal glass which has been pre-strained by a primary deformation at some point in 
the past (see the Appendix for more details on this point). 
For example, a colloidal glass subject to a shear strain below the yield strain may reasonably be 
expected to possess an anisotropic elastic response to an additional small perturbing strain 
(see (\ref{anis_elast}) for verification of this assertion). 
Taking this idea a step further, it would be of interest to investigate the nature of the yield stress surface 
in such pre-strained glasses. 
Although both the topology of the surface and its invariance with respect to hydrostatic pressure will remain 
unchanged by pre-straining, significant deviations from circularity (over and above those already arising 
from normal stresses) can be envisaged. 
In reality these deviations may well be nonstationary, decaying away as the sample ages, but such subtle dynamic 
effects are beyond current formulations of the mode-coupling theory.

\subsection*{Acknowledgements}
The authors thank M. E. Cates, M. Fuchs and Th. Voigtmann for comments on the manuscript. Funding
was
provided by the Swiss National Science Foundation.

\appendix

\subsection*{Anisotropic elasticity}
Integrating (\ref{constit1}) by parts yields the following form for the schematic constitutive 
equation
\begin{eqnarray}
\sig(t) =&& \B(t,-\infty)G(t,-\infty) - \nu_{\sigma}\one 
\notag\\
&&\hspace*{0.5cm}+
\int_{-\infty}^{t}\!\!\!dt'\,\B(t,t')\frac{\partial}{\partial t'}G(t,t').
\end{eqnarray}
For glassy states the modulus relaxes to a plateau value for long times and the contribution of the 
integral term to the overall numerical value of the stress is limited to a negligible integration 
over the $\beta$ relaxation (beyond this time the time derivative vanishes). 
The integral term may therefore be neglected to a good level of approximation. 

We first consider the situation when the system is subjected to a mixed strain field 
$\epsilonb=\epsilonb_1+\epsilonb_2$, where $\epsilonb_i=(\nabla_i\rr_i+\rr_i\nabla_i)$ is the 
infinitessimal strain due to flow $i$. 
Both of these strains are sufficiently small that the system remains in the purely elastic regime. 
The stress for long times after the application of the two strains is given by a simple 
linear superposition of the two elastic responses
\begin{eqnarray}
\sig = 
2 G_{\infty}\epsilonb_1
+ 2 G_{\infty}\epsilonb_2,
\label{lineareq}
\end{eqnarray}
as is expected from the linear theory of isotropic elasticity [Landau and Lifshitz (1986)]. 
Note that in (\ref{lineareq}) 
we have suppressed an irrelevant isotropic contribution $-\nu_{\sigma}\one$ (the system is
incompressible).

We next consider the situation whereby $\epsilonb_2$ remains small but the strain due to 
the primary flow is allowed to be sufficiently large that some plastic 
rearrangements are induced. 
We nevertheless require that the total strain must still remain sufficiently 
small that the yield stress is not exceeded and the system remains solid.
The analysis of this case neccessitates use of the partial expansion (\ref{visc4}). 
We assume that the primary strain has been applied at some time in the distant past 
and that all plastic rearrangements have ceased by the time we apply the perturbing strain 
$\epsilonb_2$.  
The Finger tensor $\B_1(t)\equiv\B_1$ is thus independent of time at the present time $t$. 
For times $t$ following application of $\epsilonb_2$
\begin{eqnarray}
\sig=\sig_1\, + \;2G_{\infty}\,\E_1\!\cdot\!\epsilonb_2\!\cdot\!\E_1^T.
\label{anisstrain}
\end{eqnarray}
By analogy with the situation considered in Sec. \ref{anisvis},
Eq.(\ref{anisstrain}) can be expressed 
in terms of a fourth rank stiffness tensor
\begin{eqnarray}
(\sig)_{ij} = (\sig_1)_{ij} 
+ 
(\C)_{ijkl}(\epsilonb_2)_{kl}
\label{anis_elast}
\end{eqnarray}
where $(\C)_{ijkl}=2G_{\infty}(\E_1)_{ik}(\E_1)_{jl}$. 
We note that the plateau value of the modulus $G_{\infty}$ may differ from its quiescent value as a result 
of plastic deformation occurring during or after application of the primary strain.

{}
\newpage

\begin{thebibliography}{}
\bibitem{footnote}
The specific numerical value for the exponent $\delta$ from the schematic model depends upon 
the path chosen in the two-dimensional space of memory function coupling constants $(\nu_1,\nu_2)$. 
By taking a standard linear path parameterized by $\varepsilon$ 
(see text below Eq.(\ref{memory})) we reproduce rather closely the viscosity divergence of the 
wavevector dependent theory for hard-spheres.
\bibitem{footnote2}
A completely analogous picture would have emerged, had we chosen a dominant shear flow 
with a perturbing compressional component. 
In this inverted thought experiment the dominant timescale would simply be set by $\gams$.
\end{thebibliography}
\end{document}